# Enabling Immersive XR Collaborations over FTTR Networks (Invited)

**Sourav Mondal and Elaine Wong**

*Department of Electrical and Electronic Engineering, Faculty of Engineering and IT*
*The University of Melbourne, Parkville, VIC 3010, Australia. sourav.mondal@unimelb.edu.au, ewon@unimelb.edu.au*

**Abstract:** Fiber-To-The-Room is a potential solution to achieve in-premise extended reality collaborations. This paper explores predictive bandwidth allocation and seamless handover schemes over FTTR, showing high-quality immersive experience for in-premise collaborations can be achieved. © 2025 The Author(s)

## 1. Introduction

Since mid-2020s, Internet technologies are evolving to support services involving human experiences by catering to the stringent requirements of immersive extended reality (XR) technologies encompassing augmented, mixed, and virtual reality (AR/MR/VR) and multi-sensory communications between humans and remote machines [1]. These technologies blend digital elements into the real world to varying extents, as shown in Fig. 1. Most are envisioned to be primarily used in in-premise domestic entertainment and industrial applications. Currently in these settings, low data rate, non-guaranteed bandwidth, and high-latency transmission technologies such as WiFi are deployed. However, to support human operators with a *comfortable quality of experience* (QoE) during XR collaborations, high-quality XR video frames (8K, 16K) are required to be transmitted with end-to-end inter-frame latency ($\leq 20$ ms) and jitter ($\leq 15$ ms) requirements [2].

## 2. Extended Reality Collaborations over FTTR

Although Beyond 5G mobile and WiFi-6/6E systems are capable of supporting standalone XR devices, they appear to be inefficient in dense in-premise scenarios. The European Telecommunications Standards Institute (ETSI) and ITU-T G.fin are considering the Fiber-To-The-Room (FTTR) and FTTR for Business (FTTR-B) where high-capacity Fiber-To-The-x (FTTx) networks are extended into the premise to be integrated with WiFi access points (WAP) [3].

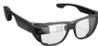

Fig. 1: Spectra of immersive XR technologies.

Fig. 2 illustrates an instance of immersive XR collaborations between a human operator and a remote machine over a FTTR-B network. The first segment is a 20 km time-division multiplexed passive optical network (TDM-PON) supporting 50 Gbps, referred to as *external PON* and the cascaded second segment is a 20 m TDM-PON supporting 10/50 Gbps, referred to as *internal PON* for deployment in a home, room, or factory. The optical line terminal (OLT) of the external PON connects multiple main FTTR units (MFs). Each MF is an optical network unit (ONU), that is also an OLT for multiple subordinate FTTR units (SFs) supporting WAPs. Data received from XR stations (STAs) at the WAPs are first buffered at the SFs and then transmitted to the MF in the next polling cycle after exchanging request and grant messages [4]. Although the latest WiFi standards have four access classes, e.g., voice, video, background, and best effort with different priority and backoff lengths, QoE of XR collaborations cannot be always gua-ranteed due to wirel-ess channel contenti-ons and uncoupled resource allocation between the fiber network and WiFi [4]. Queueing latency and uplink jitter increases, especially at high network load.

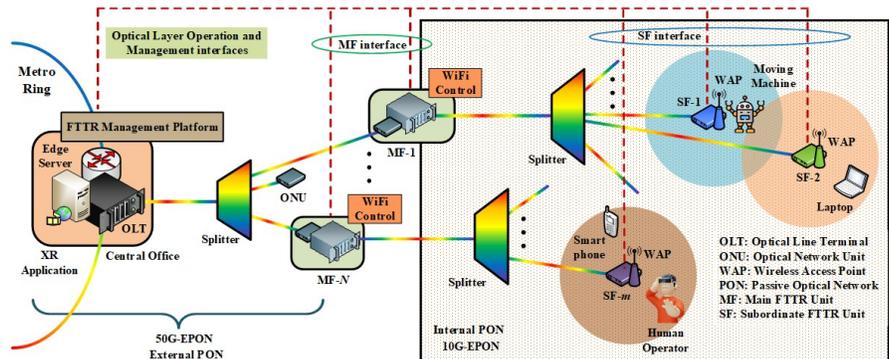

Fig. 2: Immersive XR collaborations over Fiber-To-The-Room-Business network architecture.

Another critical issue arises when the machine moves crosses the coverage areas of multiple WAPs, e.g. from SF1-to-SF2. Due to the lack of received signal power threshold-based seamless handover schemes in distributed WiFi networks, the STAs can remain attached to the same WAP for a long time despite being unable to transmit and receive any data. Addressing these challenges, we propose a *Fiber-WiFi coordinated predictive resource allocation scheme with seamless handover of XR traffic* using the control interfaces of FTTR networks.

## 3. Experimental Study

To understand the characteristics of XR traffic in a typical XR collaborations, we created a WiFi interface between a lab computer and a humanoid robot, i.e. machine, shown in Fig. 3. The robot can rotate its head-camera according to the head movements of the human. The robot's camera sends real-time video to the human. The best resolution of this camera is 640×480 with 20 frames/s and datarate ~1.152 Mbps. Furthermore, the XR frame sizes follow a truncated Gaussian distribution with a range of [50%, 150%] of the mean frame size. The inter-frame arrival times follow Gamma distribution with mean ~50 ms.

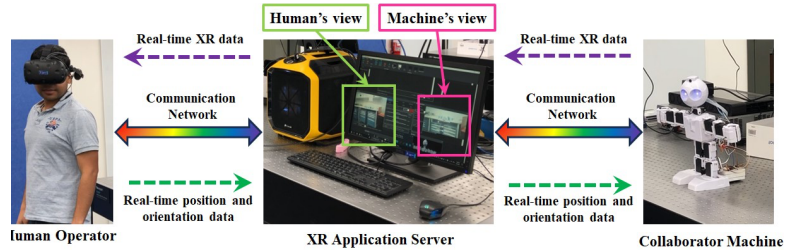

Fig. 3: Experimental study of network traffic from XR collaborations.

For our framework evaluation, the internal PON supports total 8 WiFi-6/6E WAPs maximum throughput ~9.6 Gbps. One internal PON supports 8 human operators which collaborate with 8 remote machines, supported by another internal PON. All data transmission statistics of the STA and WAPs are collected at the SFs and further relayed to the MFs and the main OLT over the control interfaces. From recent XR frames of the STAs, their future bandwidth demands and data arrival times are predicted using machine learning techniques. Accordingly, channel access to WAPs and STAs are scheduled by the *WiFi control unit* to ensure contention-free data transmissions.

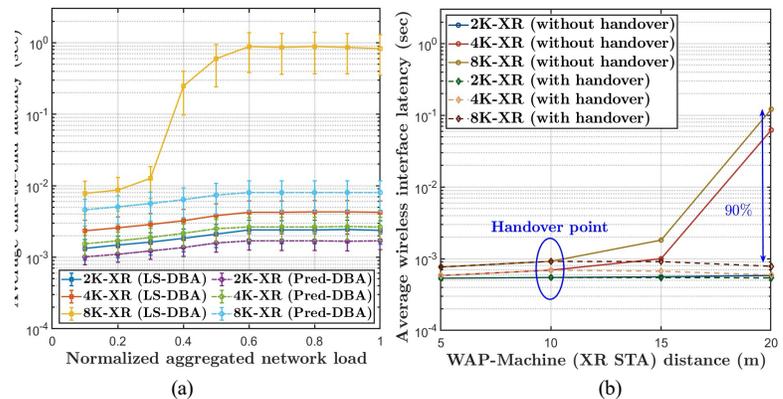

Fig. 4: (a) Average end-to-end XR inter-frame latency (ms) with LS-DBA and Pred-DBA; and (b) average wireless latency of XR traffic (ms) w/wo handover.

If the mean packet latency and jitter of an XR STA start to exceed the expected QoE requirements alongside an increase of RU demand with low-order modulation and transmit power, then a control message is transmitted to the STA to sense the received power from all neighboring WAPs. Once the XR STA sends the report, the MF can find a neighboring WAP with sufficient available bandwidth and good channel condition. Also, to ensure a seamless handover of the XR traffic, a network path between the new WAP and the XR application server is pre-established.

Fig. 4(a) plots the average end-to-end transmission latency of XR frames against normalized aggregated network load. Results show that if limited-service dynamic bandwidth allocation (LS-DBA) is used for resource allocation, then the QoE requirements are satisfied for 2K (datarate 40 Mbps) and 4K (datarate 90 Mbps) XR quality but are violated for 8K (datarate 360 Mbps) XR quality. However, with the predictive bandwidth DBA (Pred-DBA), the average end-to-end latency of XR frames are ≤15 msec under all network loads. Fig. 4(b) shows the latency reduction of XR frame transmission over the wireless interface by our proposed Fiber-WiFi framework. We consider each machine to be in a $(10\times10\times3)m^3$ room with a WiFi-6 WAP of 18 dBm transmit power. As a machine moves to a different room, an increased WAP-STA distance increases XR frame transmission latency and decreases throughput over the wireless interface, especially beyond 10 m. However, handing over at this point to a WAP with stronger signal strength reduces the 8K XR frame latency, e.g., ~90% lower than being connected with a WAP-STA distance of 20 m.